\documentclass[12pt,letterpaper]{article}

\usepackage{amsmath} 
\usepackage{amssymb} 
\usepackage{mathabx} 
\newcommand{\RNum}[1]{\uppercase\expandafter{\romannumeral #1\relax}} 

\usepackage{amsthm} 
\newtheorem{definition}{Definition} 
\newtheorem{theorem}{Theorem}[section] 
\newtheorem{lemma}[theorem]{Lemma} 
\theoremstyle{definition} 
\newtheorem{exmp}{Example} 
\newtheorem{excont}{Example} 

\usepackage{graphicx} 
\usepackage{float} 
\usepackage{subfig} 
\usepackage{booktabs} 
\usepackage{multirow} 
\usepackage{array} 
\usepackage{color} 
\usepackage{tabularx} 

\usepackage{natbib} 
\usepackage{cleveref} 

\usepackage[title]{appendix} 

\usepackage{enumitem} 
\usepackage{url} 
\usepackage{lineno} 
\usepackage{csquotes} 

\usepackage{authblk} 
\newcommand{\blind}{0}

\usepackage{lipsum} 
\usepackage[normalem]{ulem} 

\begin{document}
\def\spacingset#1{\renewcommand{\baselinestretch}%
{#1}\small\normalsize} \spacingset{1}

\newcommand{\red}[1]{\textcolor{red}{#1}}

\if0\blind
{
\title{Enumeration of regular fractional factorial designs with four-level and two-level factors}
\author[1]{Alexandre Bohyn}
\author[1]{Eric D. Schoen}
\author[1,2]{Peter Goos}
\affil[1]{Department of Biosystems, Faculty of Bioscience Engineering, KU Leuven, Leuven, Belgium}
\affil[2]{Department of Engineering Management, Faculty of Business and Economics,
University of Antwerp, Antwerp, Belgium}
\maketitle
}
\fi

\if1\blind
{
  \bigskip
  \bigskip
  \bigskip
  \begin{center}
    {\LARGE\bf Enumeration of orthogonal designs with four-level and two-level factors}
  \end{center}
  \medskip
} \fi

\bigskip
\begin{abstract}
\noindent
Designs for screening experiments usually include factors with two levels only.
Adding a few four-level factors allows for the inclusion of multi-level categorical factors or quantitative factors with possible quadratic or third-order effects.
Three examples motivated us to generate a large catalog of designs with two-level factors as well as four-level factors.
To create the catalog, we considered three methods. In the first method, we select designs using a search table, and in the second method, we use a procedure that selects candidate designs based on the properties of their projections into fewer factors. The third method is actually a benchmark method, in which we use a general orthogonal array enumeration algorithm.
We compare the efficiencies of the new methods for generating complete sets of non-isomorphic designs.
Finally, we use the most efficient method to generate a catalog of designs with up to three four-level factors and up to 20 two-level factors for run sizes 16, 32, 64, and 128.
In some cases, a complete enumeration was infeasible.
For these cases, we used a bounded enumeration strategy instead.
We demonstrate the usefulness of the catalog by revisiting the motivating examples.
\end{abstract}

\noindent%
\vfill

\newpage
\spacingset{1.45} 

\section{Introduction}
\label{sec:intro}

A major challenge in any scientific investigation is to choose an appropriate experimental design.
In this paper, we are concerned with designs for the early stages of an investigation, where there are many factors that potentially affect the responses of interest.
The  purpose in these early stages is to identify the truly influential factors, and the designs appropriate for this purpose are screening designs \citep[Chap.~1]{montgomery2017design}.
In many cases, the factors included in such designs are studied at two levels only in order to achieve run-size economy.
The most pertinent design is often selected using one or more ranking criteria.
In this regard, catalogs of designs are convenient because they offer the possibility to compare several designs on multiple criteria.
For this reason, extensive catalogs of orthogonal two-level designs have been published by \cite{chen_catalogue_1993},
\cite{block_resolution_2005}, \cite{xu_algorithmic_2009}, \cite{ryan_minimum_2010}, \cite{schoen_complete_2010} and \cite{schoen_two-level_2017}.\\

There are, however, circumstances where factors with more than two levels are involved in a screening experiment.
For example, there may be categorical factors addressing more than two categories, or numerical factors that are believed to have quadratic or cubic effects.
For these cases, designs with four-level and two-level factors, also called four-and-two-level designs, are suitable options because four-level factors can be constructed from two-level factors \citep{wu_construction_1989}.
Therefore, the run size economy of two-level designs carries over to designs with four-level as well as two-level factors.\\

All orthogonal 16-run four-and-two-level designs were enumerated by \citet{schoen_complete_2010}.
The total number of orthogonal 32-run four-and-two-level designs was too large for a complete enumeration by these authors.
However, it is possible to enumerate all regular 32-run orthogonal designs.
In such designs, any two effects in the model matrix are either completely aliased or orthogonal.
In the rest of this paper, we refer to such designs as $4^m2^{n-p}$ designs.
To construct a suitable 32-run $4^m2^{n-p}$ design, the practitioner may consult the catalogs of \citet{wu_minimum_1993} or \citet{ankenman_design_1999}.\\

Three experiments from the literature motivated us to extend the work of \citet{wu_minimum_1993} or \citet{ankenman_design_1999} to designs with more factors or larger run sizes.
\citet{schoen_designing_1999} presented two practical cases, not covered by the published catalogs.
The first one involved a 14-factor experiment on the synthesis of a catalyst on a gauze with two four-level factors.
In this experiment, a strip of gauze was cut from a roll and prepared for further processing.
The gauze was then placed in an autoclave together with a mixture of ingredients.
Next, the autoclave was heated for several hours.
After cooling, the gauze was weighted and the yield of the synthesis was determined.
The goal of the experiment was to identify the factor settings that maximize the yield of the synthesis.
The 14 factors that could affect the yield are shown in Table \ref{tab:chemical_etching_factors}.
The two-level factors 1 and 2 are related to the gauze preparation.
The four-level factors 3 and 4 as well as the two-level factors 5-10 are related to the composition of the mixture.
Finally, the two-level factors 11-14 govern the synthesis process.
A design including 32 runs was constructed based on first principles; see the original paper for more details.
The design's construction would have been much faster if a complete catalog had been available.\\

\begin{table}[htbp]
  \centering
  \caption{Factors and factor levels in the experiment on the synthesis of catalyst on a gauze}
    \begin{tabular}{llll}
    \toprule
	Stage & No. & Factor & Levels \\
	\midrule
	Gauze preparation 	& 1 & Etching 		& Yes, no \\
						& 2 & Pretreatment 	& Yes, no \\
	\midrule
	Mixture preparation & 3 	& Si source 		& 1,2,3,4 		\\
						& 4 	& Al source 		& 1,2,3,4 		\\
						& 5 	& Template 			& 1,2 			\\
						& 6 	& Si:Al ratio 		& High, low 	\\
						& 7 	& Template:Al ratio & High, low 	\\
						& 8 	& Total salt 		& High, low 	\\
						& 9 	& Water:Al ratio	& High, low 	\\
						& 10 	& Shaking time 		& Long, short 	\\
	\midrule
	Synthesis 	& 11 & Aging 				& Long, short 		\\
				& 12 & Rotating 			& Yes, no 			\\
				& 13 & Cooling 				& Forced, unforced 	\\
				& 14 & Time after synthesis & Long, short 		\\
	\bottomrule
	\end{tabular}%
  \label{tab:chemical_etching_factors}%
\end{table}%

The second example in \citet{schoen_designing_1999} is a cheese-making experiment involving 128 runs.
In this case, ten factors were investigated.
Nine factors were studied at two levels and one was studied at four levels.
The purpose of the experiment was to detect which of the factors affected the quality characteristics of the cheeses.
Here again, the design was constructed using first principles because no catalog was available; see the original paper for more details.\\

Our third example is provided by \citet{katic2011optimization}.
It is a simulation experiment about sensor fields involving 32 runs.
Two factors were studied at four levels, and five factors were studied at two levels.
\citet{wu_minimum_1993} and \citet{ankenman_design_1999} provide the same design for this case.
However, this design does not match the design used by \cite{katic2011optimization}.
In addition, as we explain later in Section \ref{sec:application}, there may be reasons to prefer yet another design, not available in the existing catalogs.
Overall, a catalog containing all the 32-run designs with two four-level factors and five two-level factors would have helped the author to consider several design options and thus make a more thoughtful design choice.\\

The purpose of our paper is to enumerate regular orthogonal designs with four-level and two-level factors for ranges of run sizes and numbers of factors that are likely to be useful in practice.
In addition, the enumeration is to be complete in the sense that all non-isomorphic designs are generated when computationally feasible, and, if not, a subset of designs with limited aliasing between main effects and two-factor interactions and among two-factor interactions is generated.
To enumerate the designs, we develop a search-table method and a method that selects candidate designs based on the properties of their projections into fewer factors.
As a benchmark method, we use the general orthogonal array enumeration algorithm of \cite{schoen_complete_2010}, which generates both regular and nonregular orthogonal designs.\\

The rest of this paper is organized as follows.
In Section \ref{sec:review}, we review the state of art in enumeration techniques for regular two-level designs.
Next, in Section \ref{sec:enum}, we adapt these methods to $4^m2^{n-p}$ designs and evaluate their computational efficiencies.
In Section \ref{sec:results}, we present a catalog of $4^m2^{n-p}$ designs for run sizes of 16, 32, 64, and 128, with up to 20 two-level factors, using the most efficient method.
In Section \ref{sec:application}, we revisit the three practical cases and explain how the catalog would have helped in the design of these experiments.
Finally, in Section \ref{sec:discussion}, we discuss our findings and the possibilities offered by our catalog.
 
\section{Enumeration of regular two-level designs}
\label{sec:review}
\subsection{Preliminaries}
\label{sec:prelim}
Consider a $2^{n-p}$ regular fractional factorial two-level design $D$, whose factor levels are coded with $-1$ and $+1$.
The design $D$ has $n$ factors and $2^{n-p}$ runs.
The $n$ factors include $k=n-p$ basic factors whose $2^k$ level combinations are all present in the design.
There are also $p$ added factors, defined by interactions between the $k$ basic factors.
These interactions are called generators.
The element-wise product between an added factor and its generator results in the identity column, which is a column whose elements are all $+1$.
If we represent the factors as lower case letters, then such a product can be represented as a word containing the letter representing the added factor and the letters corresponding to the basic factors involved in the generator.
All words thus correspond to the identity column, as does any product of two or more words.
All words derived from the $p$ generators together with all words formed by the products of two or more of them form the defining relation.
Therefore, the defining relation contains $2^{p}-1$ words.

\begin{exmp}
\label{exmp:two-lvl-des}
Let $D$ be a $2^{6-2}$ design with four basic factors $a,b,c,d$ and two added factors defined as $e=abc$ and $f=acd$.
The products between the generators, $abc$ and $acd$, and the added factors, $e$ and $f$, result in the words $abce$ and $acdf$, respectively. Multiplying these words results in the word $abceacdf$ which can be simplified to $bdef$ because the product of a factor with itself results in the identity column.
Therefore, the defining relation of $D$ is the set of $2^{2}-1$ words

\begin{equation}
\{abce, acdf, bdef\}\text{.}
\label{eq:ex_design}
\end{equation}
\qed
\end{exmp}

The length of a word is the number of letters it contains.
For a $2^{n-p}$ design $D$, let $A_{i}$ denote the number of words of length $i$ in the defining relation.
The vector
\begin{equation}
    W(D) = \left(A_3,A_4,\ldots,A_{n}\right)
\end{equation}
is called the word length pattern (WLP) of the design.
A word of length three indicates a subset of three factors whose main effects are fully aliased with the interaction between the other two factors.
A word of length four indicates a subset of four factors where the two-factor interaction of any two of the factors is fully aliased with the two-factor interaction of the two remaining factors.
Words of length five or higher indicate less serious aliasing, such as aliasing between a two-factor interaction and a three-factor interaction.\\

The resolution of a $2^{n-p}$ design $D$ is the smallest integer $r$ for which $A_{r}>0$, and thus the length of the shortest word in the defining relation \citep{box_2_1961}.
To distinguish between designs with the same resolution, \cite{fries_minimum_1980} proposed the aberration criterion:

\begin{definition}
For any two $2^{n-p}$ designs $D_{1}$ and $D_{2}$, let $r$ be the smallest integer for which $A_{r}\left(D_{1}\right) \neq A_{r}\left(D_{2}\right)$.
Then, $D_{1}$ has less aberration than $D_{2}$ if $A_{r}\left(D_{1}\right)<A_{r}\left(D_{2}\right)$.
If there is no design with less aberration than $D_{1}$, then $D_{1}$ has minimum aberration (MA).
\end{definition}

A $2^{n-p}$ design can be represented as an $2^{n-p} \times n$ array where each row is an experimental run and each column is a factor.
The level of factor $j$ in the $i$th experimental run is determined by the value in position $(i,j)$ of the array.
The levels are denoted by $-1$ and $+1$, representing the low and high level of a quantitative two-level factor, or the two categories of a categorical factor.\\

Two designs are isomorphic if one can be obtained from the other through permutations of rows, columns and levels within a given column.
If this is the case, there exists an isomorphic map from one design to the other, denoted by $\pi \equiv (\kappa, \rho, \sigma)$, that is a specific sequence $\kappa$ of column permutations, a specific sequence $\rho$ of row permutations and a specific set $\sigma$ of level switches of the columns.
It is easy to see that two isomorphic designs have the same word length pattern.
Designs that are isomorphic to each other belong to the same isomorphism class.
As isomorphic designs have exactly the same statistical properties, it suffices to enumerate one design per isomorphism class.

\begin{excont}[continued] 
The word length pattern $W$ of design $D$ is $W\left(D\right) = \left(0,3,0,0\right)$.
Table 2 of \cite{chen_catalogue_1993} shows four regular six-factor 16-run designs, each with a different word length pattern.
Any regular $2^{6-2}$ design has to be isomorphic to one of these four designs.
The design $D$ is isomorphic to design 6-2.1 in Table 2 of \cite{chen_catalogue_1993} and therefore has MA.
\qed
\end{excont}

\subsection{Enumeration procedures}
\label{sec:enumeration}
A minimal complete set (MCS) of $2^{n-p}$ designs of resolution $R$, denoted by $C^{R}_{n,p}$, is a set of designs with exactly one representative of each isomorphism class. 
A complete set of $2^{n-p}$ designs of resolution $R$, denoted by $\widetilde{C}^{R}_{n,p}$, is a set of designs with at least one representative for each isomorphism class.
The most common method to generate a MCS of $2^{(n+1)-(p+1)}$ designs of resolution $R$ is to first extend $C^{R}_{n,p}$, a MCS of $n$-factor parent designs, into $\widetilde{C}^{R}_{n+1,p+1}$, a complete set of $(n+1)$-factor candidate designs, and then reduce it to a MCS $C^{R}_{n+1,p+1}$.
This procedure can be divided in two main steps:
\begin{description}
	\item[Extension:] Each parent design in $C^{R}_{n,p}$ is extended into one or more $2^{(n+1)-(p+1)}$ candidate designs by adding candidate two-level factor columns so as to form $\widetilde{C}^{R}_{n+1,p+1}$.
	\item[Reduction:] Of all the candidate designs in $\widetilde{C}^{R}_{n+1,p+1}$, only one representative per isomorphism class is retained.
\end{description}
We now discuss the extension and reduction methods that have appeared in the literature.

\subsubsection{Extension methods}
\label{sec:extension}

\paragraph{Search table}
\cite{bingham_minimum-aberration_1999} used a search table \citep{franklin_selection_1977} to enumerate complete sets of two-level split-plot designs.
Applied to a $2^{n-p}$ design with $2^k$ runs, $k=n-p$ basic factors and $p$ added factors, without split-plot complication, a search table is a two-way table with ${2^{k}-k-1}$ rows and $p$ columns.
The columns represent the added factors, while the rows correspond to all possible interactions between two or more basic factors.
The interactions are the possible generators for the added factors.
In the table, the interactions are sorted according to their order.
Interactions of the same order are sorted lexicographically.
 
\begin{excont}[continued]
The search table for $D$ is presented in Table \ref{tab:searchtable_2lvl}.
For the first added factor $e$, there are $2^{4}-4-1=11$ possible generators.
However, generators of the same order result in isomorphic 5-factor designs, so that there are only three isomorphism classes. 
The words $abe$, $abce$ and $abcde$ are used to represent these classes because they appear first in the table.
These three $2^{5-1}$ non-isomorphic designs are the ones presented in Table 2 of \citet{chen_catalogue_1993}.\\

For the second added factor $f$, \cite{bingham_minimum-aberration_1999} explained that we only need to consider generators that are lower in the table than the generator used for the first added factor.
Indeed, using $ab$ as generator both for factors $e$ and $f$ would result in a word of length two $(ef)$ indicating that the main effects of factors $e$ and $f$ are aliased.
The design would thus have a resolution of \RNum{2} which is undesirable.
Therefore, the choice of an additional generator in the same row has to be disregarded.
Next, consider the pair of words $\{abce, acf\}$ where the generator of the second word, $ac$, appears higher in the table than the generator of the first word, $abc$.
The resulting design can be obtained from another pair of words, for which the generator of the second word appears lower in the search table than the generator of the first word.
This can be seen by interchanging the columns and relabelling the factors in the original pair of words in the following way: $\{abce,acf\} \rightarrow \{acbe,abf\} \rightarrow \{acbf, abe\} \rightarrow \{abe, abcf\}$.
This also applies to any of the other sets of words containing $abce$ or $abcde$ and an additional word from a higher row in the table.
We conclude that we do not lose non-isomorphic designs if we only consider generators that appear lower in the search table.
The words that are not considered for the generation of this $2^{6-2}$ design are struck through.

Without the search table, each of the three non-isomorphic $2^{5-1}$ designs, represented by the words $abe$, $abce$ and $abcde$, can generate 11 candidates, one for each generator available.
With the search table, the design with $abe$ as its first word can only generate 10 candidates, the design with $abce$ as its first word can only generate 4 candidates and the design with $abcde$ as its first word cannot generate any candidates.
As a consequence, the search-table approach leads to 14 $2^{6-2}$ candidate  designs among the 33 possible options.
These 14 options must belong to one of the four isomorphism classes identified by \citet{chen_catalogue_1993}.
\qed
\end{excont}

\begin{table}[hbtp]
        \centering
        \caption{Search table for $2^{6-2}$ designs where the three columns for factor $f$ correspond with the possible generators for factor $e$, written in parentheses.}
        \begin{tabular}{lcccc}
        \toprule
        \multicolumn{1}{l}{\multirow{2}[4]{*}{{\textbf{Generator}}}} & \multicolumn{4}{c}{\textbf{Added factor}} \\
    	\cmidrule{2-5}	&	$e$	&	$f$ $(ab)$ & $f$ $(abc)$ & $f$ $(abcd)$\\
    	\midrule
        $ab$	& $abe$ 			& \sout{$abf$} 	& \sout{$abf$} 	& \sout{$abf$} 	\\
        $ac$	& \sout{$ace$} 		& $acf$ 		& \sout{$acf$}	& \sout{$acf$}	\\
        $ad$	& \sout{$ade$} 		& $adf$ 		& \sout{$adf$}	& \sout{$adf$}	\\
        $bc$	& \sout{$bce$} 		& $bcf$ 		& \sout{$bcf$}	& \sout{$bcf$}	\\
        $bd$	& \sout{$bde$} 		& $bdf$ 		& \sout{$bdf$}	& \sout{$bdf$}	\\
        $cd$	& \sout{$cde$} 		& $cdf$ 		& \sout{$cdf$}	& \sout{$cdf$}	\\
        $abc$ 	& $abce$ 			& $abcf$ 		& \sout{$abcf$} & \sout{$abcf$}	\\
        $abd$ 	& \sout{$abde$} 	& $abdf$ 		& $abdf$ 		& \sout{$abdf$}	\\
        $acd$ 	& \sout{$acde$} 	& $acdf$ 		& $acdf$ 		& \sout{$acdf$}	\\
        $bcd$ 	& \sout{$bcde$} 	& $bcdf$ 		& $bcdf$ 		& \sout{$bcdf$}	\\
        $abcd$ 	& $abcde$        	& $abcdf$ 		& $abcdf$ 		& \sout{$abcdf$}\\
        \bottomrule
        \end{tabular}%
        \label{tab:searchtable_2lvl}%
    \end{table}%

\paragraph{Delete-one-factor projection}
For a $2^{(n+1)-(p+1)}$ design $D_{c}$ and $i=1,\ldots,n+1$, let $D_{c(i)}$ be the $2^{n-p}$ subdesign obtained by deleting the $i$th column.
Such a subdesign is called a delete-one-factor projection (DOP) of $D_{c}$ \citep{xu_algorithmic_2009}.
For any design $D$ in $C^{R}_{n,p}$, any interaction of the basic factors that has not yet been used to construct the design is a candidate column, and using any one of the candidate columns for a new added factor yields a $2^{(n+1)-(p+1)}$ candidate $D_c$.
\cite{xu_algorithmic_2009} showed that discarding $D_c$ if its resolution is lower than $R$ or if $D$ does not have MA among all the DOPs of $D_c$ still results in a complete set  $\widetilde{C}^{R}_{n+1,p+1}$.
This is because all non-isomorphic $n$-factor designs are available for extension, including a design that is isomorphic to the MA DOP of $D_c$.

\paragraph{Minimum complete set algorithm}
The minimum complete set (MCS) algorithm was introduced by \citet{schoen_complete_2010} to generate MCSs of orthogonal arrays of specific run sizes, numbers of factor levels and strengths.
The algorithm uses the lexicographically minimum in columns (LMC) form:

\begin{definition}
Consider two $2^{n-p}$ designs $D_{1}$ and $D_{2}$ with $N$ runs.
Denote by $d_{1}$ and $d_{2}$ the $N \cdot n$ vectors obtained by concatenating the $n$ columns of $D_{1}$ and $D_{2}$, respectively, and by $d_{i j}(i=1,2 ; j=1, \ldots ,N\cdot n)$ the elements of these vectors.
Then, $D_{1}$ is lexicographically smaller than $D_{2}$ if there is an $l \leq N \cdot n$ such that $d_{1 j}=d_{2 j}$ for $j=1, \ldots ,l-1$ and $d_{1 l}<d_{2 l}$.
The design $D_{1}$ is in LMC form if no other design from its isomorphism class is lexicographically smaller.
\end{definition}

The MCS algorithm starts with a MCS of parents, $C^{R}_{n,p}$, that are all in LMC form.
Then, each parent is extended to a candidate design by creating a new column using element-wise addition.
Only additional columns that are lexicographically smaller than the columns of the parent are considered.
The set of extended designs is guaranteed to contain all designs from $C^{R}_{n+1,p+1}$ in LMC form \citep{schoen_complete_2010}.
However, the MCS algorithm generates both regular and non-regular designs, that is, designs where the aliasing between the effects in the model matrix can also be partial.
For example, from the three non-isomorphic $2^{5-1}$ designs from Example \ref{exmp:two-lvl-des}, corresponding to words $abe$, $abce$ and $abcde$, the LMC algorithm produces three non-regular designs and four regular $2^{6-2}$ designs.

\subsubsection{Reduction methods}
\label{sec:reduction}
\paragraph{Partitioning}
The reduction step is a computationally intensive part of the whole enumeration process.
For this reason, it is often preceded by a partitioning step in which a design criterion is selected such that isomorphic designs have the same value and any two designs with different values must be non-isomorphic.
Such a criterion is called an invariant.
First, an invariant is computed for all candidates in $\widetilde{C}^{R}_{n+1,p+1}$.
Then, the set of candidates is divided into subsets of designs with the same value for the invariant.
Any singleton is automatically an isomorphism class representative and does not need to go through the reduction step.\\

\cite{chen_catalogue_1993} used a letter pattern in addition to the WLP as an invariant in their algorithm, while \cite{lin_isomorphism_2008} used the eigenvalues of a word pattern matrix in addition to the WLP.
\cite{xu_algorithmic_2009} used the moment projection pattern \citep{xu_moment_2005}, a ranking criterion based on the Hamming distance of the projections of the design matrix, as an invariant in his algorithm.

\paragraph{Pairwise isomorphism testing}
\Citet{chen_catalogue_1993} implemented a pairwise isomorphism testing procedure where, within every subset of $\widetilde{C}^{R}_{n+1,p+1}$ created by the partitioning, each candidate design is tested against all other candidates.
For each pair of candidates, every possible isomorphic map from one design to the other is tested and if none can be found, the pair of designs are non-isomorphic.

\paragraph{Canonical form rejection}
\cite{schoen_complete_2010} employed lexicographic ordering in their MCS algorithm.
All candidate designs generated in the extension step of the algorithm are tested to determine whether they are in LMC form.
If that is the case, the candidate is kept as an isomorphism class representative.
Otherwise, it is rejected.
However, this procedure also retains non-regular designs.
Since we address only regular designs, we perform an additional regularity check for every candidate design $D$.
Let $\mathbf{X}$ be the $N \times \left(2^{k}-1\right)$ model matrix corresponding to the interactions between the $k$ basic factors of the design $D$ and let $\mathbf{D}$ be its $N \times n$ design matrix.
Then, the design is regular and therefore retained if and only if the $\left(2^{k}-1\right) \times n$ matrix $ N^{-1}\mathbf{X}^{\prime}\mathbf{D}$ only contains ones and zeros.

\paragraph{Canonical form conversion}
NAUTY \citep{mckay_practical_2014} is a program that selects isomorphism class representatives among sets of graphs.
It converts all graphs in the set into a specific form called the NAUTY canonical form and then selects one representative for each unique form.
\citet{ryan_minimum_2010} showed that a $2^{n-p}$ design can be uniquely represented as a graph, and that graph isomorphism is equivalent to design isomorphism.
This justifies their use of NAUTY to perform the isomorphism reduction step.

\subsubsection{State of the art in two-level design enumeration}
\cite{chen_catalogue_1993} were the first to apply extension and reduction to enumerate two-level regular designs.	
They considered all unused two-level factor interaction columns in the extension step, and they used a pairwise isomorphism testing in the reduction step.
Their catalog includes all non-isomorphic 16-run designs of resolution \RNum{3}, resolution-\RNum{3} 32-run designs with up to 28 factors, and resolution-\RNum{4} 64-run designs with up to 32 factors.
\citet{block_resolution_2005} extended that catalog to 128-run designs of resolution \RNum{4} with up to 64 factors by differentiating design based on their projections rather than performing a pairwise isomorphism check.
\cite{xu_algorithmic_2009} used the moment projection pattern criterion for partitioning, pairwise isomorphism testing and the DOP method in the extension step of the algorithm, thereby creating a more efficient algorithm.
This allowed him to enumerate and present efficient designs for run sizes up to 4096, with resolution up to \RNum{7}. 
\cite{ryan_minimum_2010} improved the speed of Xu's algorithm by replacing the pairwise isomorphism testing procedure in the reduction step with the NAUTY-based algorithm.
With this procedure, they extended the catalog of \cite{xu_algorithmic_2009} by enumerating resolution \RNum{4} 128-run and 256-run designs with up to 64 factors, and addressing further resolution-\RNum{6} cases for designs involving 2048 and 4096 runs.

\section{Enumeration of $4^{m}2^{n-p}$ designs}
\label{sec:enum}
\subsection{Preliminaries}
\label{sec:fatl_prelim}
A $4^m2^{n-p}$ design has $m$ four-level factors and $n$ two-level factors.
Using the grouping scheme of \cite{wu_construction_1989}, these designs can be derived from two-level designs by combining pairs of two-level factors into four-level factors. 
Table \ref{tab:grouping_scheme} illustrates this by showing that each of the four level combinations of two two-level factors can be assigned to a different level of a four-level factor.
The two-level factors constitute two parts of the main effect of the four-level factor.
The third part is the interaction between the two two-level factors.
The two main effects and the interaction are called pseudo-factors.\\

\begin{table}[ht]
\centering
\caption{Grouping scheme \citep{wu_construction_1989} to combine two two-level factors, $a$ and $b$, into a four-level factor, $A$.} 
\begin{tabular}{cccc}
		\toprule
        $a$ 	& $b$ 	& 				& $A$ \\
        \midrule
        $+1$ 	& $+1$ 	& $\rightarrow$  	& 0 \\
        $+1$ 	& $-1$ 	& $\rightarrow$  	& 1 \\
        $-1$ 	& $+1$ 	& $\rightarrow$ 	& 2 \\
        $-1$ 	& $-1$ 	& $\rightarrow$ 	& 3 \\
        \bottomrule
\end{tabular}
\label{tab:grouping_scheme}
\end{table}

A regular $4^{m}2^{n-p}$ design is constructed from $k$ two-level basic factors such that it has $2^k$ runs. 
To define $m$ four-level factors, $2m$ of the $k$ basic factors are used. 
The remaining $k-2m$ basic factors serve as building blocks for the $p = 2m + n - k$ additional two-level factors.
A MCS of $4^m2^{n-p}$ designs involving $2^k$ runs with resolution $R$ is denoted by $C^{R}_{m,n,p}$, while a complete set is denoted by $\widetilde{C}^{R}_{m,n,p}$.

To define the word length pattern of a $4^{m}2^{n-p}$ design, we need to adjust the definition of the word length pattern of two-level designs.
More specifically, an interaction of two pseudo-factors that form a four-level factor constitutes another pseudo-factor.
For this reason, the length of a word that contains two pseudo-factors corresponding to a single four-level factor is decreased by 1.
Following \cite{wu_minimum_1993}, we represent the three pseudo-factors corresponding to a single four-level factor as lower case letters with an index $i \in \{1,2,3\}$.
For instance, when the original factors $a$ and $b$ define the four-level factor $A$, then we replace the factors $a$ and $b$ with $a_1$ and $a_2$, and their product $ab$ by $a_3$ in all the words of the defining relation of the design.

\begin{exmp}
Consider the $2^{6-2}$ design $D$ with a defining relation $\{abce,acdf,bdef\}$ presented in Example \ref{exmp:two-lvl-des}.
We turn it into a $4^12^{4-2}$ design, $D_1$, by creating the four-level factor $A$ using the two two-level factors $a$ and $b$ and their interaction $ab$ as pseudo factors.
By relabeling the pseudo-factors, the three words of the original defining relation $\{abce,acdf,bdef\}$ become $\{a_3ce,a_1cdf, a_2def\}$, and now have lengths 3, 4 and 4, respectively.
The word length pattern of the design therefore changes from $W\left(D\right) = (0,3,0,0)$ to $W\left(D_{1}\right) = (1,2,0,0)$, and $D_1$ becomes a resolution-\RNum{3} design instead of a resolution-\RNum{4} design.
Furthermore, the two-factor interaction $ce$ is now fully aliased with one of the main effects of the four-level factor $A$.
\qed
\end{exmp}

\citet{wu_minimum_1993} differentiate between words containing pseudo-factors and words not containing any pseudo-factors.
Words involving pseudo-factors from $t$ different four-level factors are of type $t$ and words not involving pseudo-factors are of type 0.
This distinction helps to differentiate non-isomorphic designs according to the interest of the practitioner.
For example, a design whose only words are of type 0 is especially useful to study the main effects and two-factor interactions of four-level factors, because these effects are not aliased with effects of two-level factors or with effects of other four-level factors.
Similarly, a design with type-2 words of length three has main effects of a two-level factor that are aliased with an interaction between two four-level factors, so that the design is less suitable to study these two-level factors.\\

For a $4^m2^{n-p}$ design $D$, let $A_{it}$ denote the number of words of length $i$ and type $t$, and let $\mathbf{A}^{m}_{i}=(A_{im},\ldots,A_{i0})$, be a vector representing all words of length $i$ in descending order of the type.
We call the vector
\begin{equation}
\mathbf{W}_{m}(D) = \left(\mathbf{A}^{m}_{3},\ldots,\mathbf{A}^{m}_{m+n}\right)
\end{equation}
the word length pattern of type $m$, abbreviated as WLP$_{m}$.
In a similar fashion, let $\mathbf{A}^{0}_{i}=(A_{i0},\ldots,A_{im})$ be a vector representing all words of length $i$ in ascending order of the type.
We call 
\begin{equation}
    \mathbf{W}_{0}(D) = \left(\mathbf{A}^{0}_{3},\ldots,\mathbf{A}^{0}_{m+n}\right)
\end{equation}
the word length pattern of type 0, abbreviated as WLP$_{0}$.
If $m>1$, the vector $\mathbf{A}_{i}$ includes three or more elements so that there are at least six ways to order them.
We find it hard to motivate orderings of the $A_{it}$ values other than the descending or ascending orders of $t$.
For this reason, we only consider WLP$_{m}$ and WLP$_{0}$.
The WLP$_{m}$ should be especially useful as a design criterion if the emphasis is on studying the effects involving the four-level factors.
In contrast, the WLP$_{0}$ should be useful if the emphasis is on studying the two-level factors.\\

The resolution of a $4^{m}2^{n-p}$ design $D$ is defined as the smallest integer $r$ for which $\mathbf{A}^{0}_r$ or $\mathbf{A}^{m}_r$ is a non-zero vector, and thus as the length of the shortest word in the defining relation.
However, to define aberration, we need to differentiate the words of different types.

\begin{definition}[\cite{wu_minimum_1993}]
Consider two $4^m2^{n-p}$ designs $D_1$ and $D_2$.
Let $\mathbf{W}_{t}(D_1)$ and $\mathbf{W}_{t}(D_2)$ be the word length patterns of type $t$ of $D_1$ and $D_2$, respectively.
Then, $D_1$ has less aberration of type $t$ than $D_2$ if the first entry where $\mathbf{W}_{t}(D_1)$ differs from $\mathbf{W}_{t}(D_2)$ is smaller in $\mathbf{W}_{t}(D_1)$ than in $\mathbf{W}_{t}(D_2)$.
If no other design has less aberration of type $t$ than $D_1$, then $D_1$ has minimum aberration of type $t$.
\end{definition}

\begin{excont}[continued]
We return to the $4^12^{4-2}$ design $D_1$.
The two-level basic factors are $c$ and $d$, and there are two added two-level factors $e=abc$ and $f=acd$.
The original defining relation of $D_1$ was $\{abce,acdf,bdef\}$ and became $\{a_3ce,a_1cdf, a_2def\}$ after relabeling the pseudo-factors corresponding to the four-level factor.
In the new defining relation, one word has length 3 and two words have length 4, but they are all words of type 1 since they all contain one pseudo-factor.
Therefore, we can say that $A_{31}=1$ and $A_{41}=2$, and thus the word length pattern of type $1$ of $D_1$ is $\mathbf{W}_{1}(D_1)=\left((1,0),(2,0)\right)$, while its word length pattern of type $0$ is $\mathbf{W}_{0}(D_1)=\left((0,1),(0,2)\right)$.
According to \citet{wu_minimum_1993}, the MA $4^12^{4-2}$ design of both type $1$ and 0 has a defining relation $\{acde,bcf,abdef\}$, which becomes $\{a_1cde,a_2cf,a_3def\}$ after relabeling the pseudo-factors corresponding to the four-level factor, and it has a word length pattern of type 1 of $\left((1,0),(2,0)\right)$, and a word length pattern of type 0 of $\left((0,1),(0,2)\right)$. 
We see that by performing the relabeling $e\leftrightarrow f$, and $a_1 \rightarrow a_3 \rightarrow a_2 \rightarrow a_1$, the two designs have the same word length patterns.
Therefore, $D_1$ is a MA design of type 1 and of type 0.
\qed
\end{excont}

\subsection{Extension procedures}
\label{sec:two_level_adpat}
\paragraph{Search table}
The search table has to be adapted to cope with $4^{m}2^{n-p}$ designs.
More specifically, we need to relabel the pseudo-factors corresponding to the four-level factors.
Just as for two-level designs, the columns of the search table represent the added factors, while the rows correspond to all possible interactions between two or more basic factors.
These interactions are the possible generators for the added factors.
We first relabel the interactions with the pseudo-factors corresponding to the four-level factors.
Then, we sort the interactions by their order.
However, pseudo-factors cannot be relabeled to ordinary factors, and ordinary factors cannot be relabeled to pseudo-factors.
Interactions with the same order are further sorted by type.
Interactions with the same length and type are sorted lexicographically.
After the search table has been sorted correctly, the same rules as for two-level designs apply.
This means that the generators considered for an additional added factor must not come from the same row or a higher row than the previous one added.\\

We construct $4^m 2^{n-p}$ designs in such a way that the sub-designs involving only the four-level factors are full factorial designs which are possibly replicated.
The main reason for this measure is that it simplifies the construction of search tables, because it permits the inclusion of the four-level factors as basic factors.
In addition, it implies that there is no aliasing among the four-level factors.
So designs with a fractional sub-design of the four-level factors are disregarded.\\

\begin{excont}[continued]
We return to the $4^12^{4-2}$ design $D_1$.
The two-level basic factors are $c$ and $d$, and there are two added two-level factors $e=abc$ and $f=acd$. 
The adapted search table for $D_1$ is presented in Table \ref{tab:searchtable_mixedlvl}.
To visualize the relabeling, the original generators are listed in the first column of the table, and to visualize the new way of sorting, the order and type of the generators are also indicated in the search table.
It is easy to see that $a_3$ cannot be considered as a generator since it has order 1.
This would imply aliasing between the main effect of the four-level factor and an added two-level factor.
We also see that, after relabeling, $ac$, $bc$ and $abc$ become $a_1c$, $a_2c$ and $a_3c$, which are all generators of the same order and of the same type.
Since pseudo-factors can only be relabeled to pseudo-factors, and two-level factors to two-level factors, the designs with $cde$ and $a_1ce$ as the first added factor are not isomorphic to each other.
However, all generators with order 2 and type 1 will lead to isomorphic designs.
All other generators with order 3 and type 1 will also lead to isomorphic designs.
Therefore, there are three non-isomorphic $4^12^{3-1}$ designs with 16 runs, with $cde$, $a_1ce$ and $a_1cde$ as the first added factor.
In $D_1$, the original generator of factor $e$ is $a_3c$, but the generator $a_1c$ results in a $4^12^{3-1}$ design that is isomorphic to the original one.
The generators that lead to isomorphic designs are struck through.
Without the search table, each of the three non-isomorphic $4^12^{3-1}$ designs could generate ten candidates, one for each generator with order two or more.
With the search table, the design with $cde$ as its first word can generate nine candidates, the design with $a_1ce$ as its first word can generate eight candidates and the design with $a_1cde$ as its first word can only generate two candidates, since it is located near the bottom of the search table.
As a consequence, the search-table approach leads to 19 $4^12^{4-2}$ candidate  designs among the 30 possible options.
\qed
\end{excont}

\begin{table}[ht]
    \centering
    \caption{Search table for $4^{1}2^{4-2}$ designs with $A=(a,b,ab)$ as four-level factor where the three columns for factor $f$ correspond with the possible generator for factor $e$, written in parentheses.}
    \begin{tabular}{llcccccc}
        \toprule
        \multicolumn{1}{c}{\multirow{2}[3]{2cm}{Original generator}} & \multicolumn{1}{c}{\multirow{2}[3]{2cm}{Relabeled generator}} & \multicolumn{1}{c}{\multirow{2}[3]{*}{Order}} & \multicolumn{1}{c}{\multirow{2}[3]{*}{Type}} & \multicolumn{4}{c}{Added factor} \\
        \cmidrule{5-8}  &       &       &       & $e$               & $f$ $(cd)$    & $f$ $(a_1c)$ & $f$ $(a_1cd)$ \\
        \midrule
        $ab$		& $a_3$		& 1     & 1     & \sout{$a_3e$}	    & \sout{$a_3f$} & \sout{$a_3f$} & \sout{$a_3f$} \\
        $cd$ 		& $cd$ 		& 2 	& 0 	& $cde$ 		    & \sout{$cdf$}	& \sout{$cdf$}	& \sout{$cdf$}	\\
        $ac$ 		& $a_1c$ 	& 2 	& 1 	& $a_1ce$ 		    & $a_1cf$ 	    & \sout{$a_1cf$}& \sout{$a_1cf$}	    \\
        $bc$ 		& $a_2c$ 	& 2 	& 1 	& \sout{$a_2ce$}    & $a_2cf$ 	    & $a_2cf$ 	    & \sout{$a_2cf$}	    \\
        $abc$ 		& $a_3c$ 	& 2 	& 1 	& \sout{$a_3ce$}    & $a_3cf$ 	    & $a_3cf$ 	    & \sout{$a_3cf$}	    \\
        $ad$ 		& $a_1d$ 	& 2 	& 1 	& \sout{$a_1de$}    & $a_1df$ 	    & $a_1df$ 	    & \sout{$a_1df$}	    \\
        $bd$ 		& $a_2d$ 	& 2 	& 1 	& \sout{$a_2de$}    & $a_2df$ 	    & $a_2df$ 	    & \sout{$a_2df$}	    \\
        $abd$ 		& $a_3d$ 	& 2 	& 1 	& \sout{$a_3de$}    & $a_3df$ 	    & $a_3df$ 	    & \sout{$a_3df$}	    \\
        $acd$ 		& $a_1cd$ 	& 3 	& 1 	& $a_1cde$ 		    & $a_1cdf$ 	    & $a_1cdf$ 	    & \sout{$a_1cdf$} 	    \\
        $bcd$ 		& $a_2cd$ 	& 3 	& 1 	& \sout{$a_2cde$}   & $a_2cdf$ 	    & $a_2cdf$ 	    & $a_2cdf$ 	    \\
        $abcd$ 		& $a_3cd$ 	& 3 	& 1 	& \sout{$a_3cde$}   & $a_3cdf$ 	    & $a_3cdf$ 	    & $a_3cdf$ 	    \\
        \bottomrule
    \end{tabular}
    \label{tab:searchtable_mixedlvl}%
\end{table}

\paragraph{Delete-one-factor projections}
The DOP method can be adapted to $4^m2^{n-p}$ designs by only considering the deletion of two-level factors.
For any design $D$ in $C^{R}_{m,n,p}$, adding a two-level column yields a $4^{m}2^{(n+1)-(p+1)}$ candidate design $D_c$. 
Such a candidate is discarded if its resolution is lower than $R$ or if $D$ does not have MA among all DOPs of $D_c$.
The resulting set, $\widetilde{C}^{R}_{m,n+1,p+1}$, is a complete set.
The proof can be found in Appendix~\ref{appendix:proofs}.

\subsection{Selected procedures}
\label{sec:selected_procedures}
For the enumeration of $4^m2^{n-p}$ designs, we compare three methods.
First, we modify the method of \cite{ryan_minimum_2010}, such that the adapted DOP algorithm from Section \ref{sec:two_level_adpat} performs the extension step, and the NAUTY-based algorithm performs the reduction step.
We call this method \enquote{DOP-NAUTY}.
However, the DOP method requires many WLP computations, and
\cite{xu_generalized_2001} showed that the WLP computation for mixed-level designs (such as $4^m2^{n-p}$ designs) is more complex than for pure-level designs (such as $2^n$ designs).
In contrast, the search-table method does not involve WLP computations, because it considers candidates based on a fixed search table.
Therefore, it might be faster than the DOP method.
For this reason, we also consider a method using the search-table algorithm in the extension step and NAUTY to check isomorphism.
We call it \enquote{ST-NAUTY}.
The third method we consider is a more general method that uses the MCS algorithm of \cite{schoen_complete_2010} for the extension step and retains the regular non-isomorphic designs in the reduction step.
We call that third method \enquote{MCS-regular}.
For reproducibility purposes, pseudo-codes for the ST-NAUTY and DOP-NAUTY methods are available in the supplementary materials of the paper.
Pseudo-code for the benchmark MCS method is available in \cite{schoen_complete_2010}.
The code used for the enumeration of the catalog is also available on Github at \url{https://github.com/ABohynDOE/enumeration_fatld}.

\section{Enumeration results}
\label{sec:results}

\subsection{Computing times}
\label{sec:computing_time}
\begin{figure}[hbtp]
\centering
\subfloat[32-run resolution 
\RNum{3} $4^12^{n-p}$ designs]{%
  \includegraphics[clip,height=0.25\textheight]{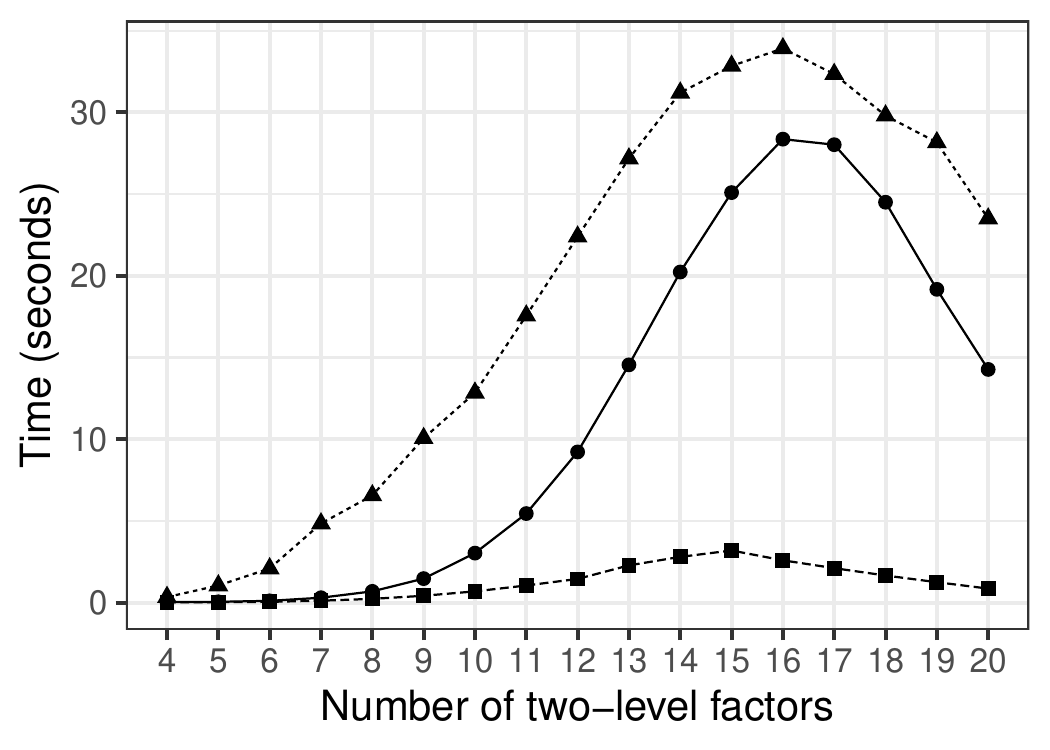}%
  \label{fig:test_1}%
  }

\subfloat[64-run resolution 
\RNum{4} $4^12^{n-p}$designs]{%
  \includegraphics[clip,height=0.25\textheight]{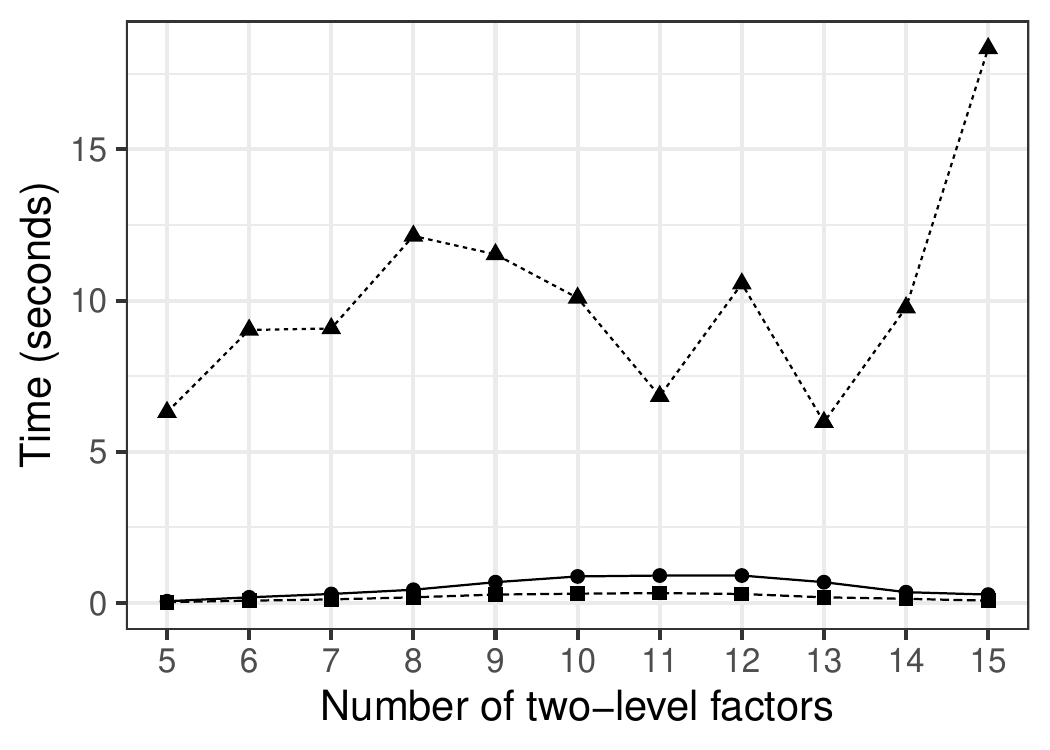}%
  \label{fig:test_2}%
}

\subfloat[64-run resolution 
\RNum{4} $4^22^{n-p}$ designs]{%
  \includegraphics[clip,height=0.25\textheight]{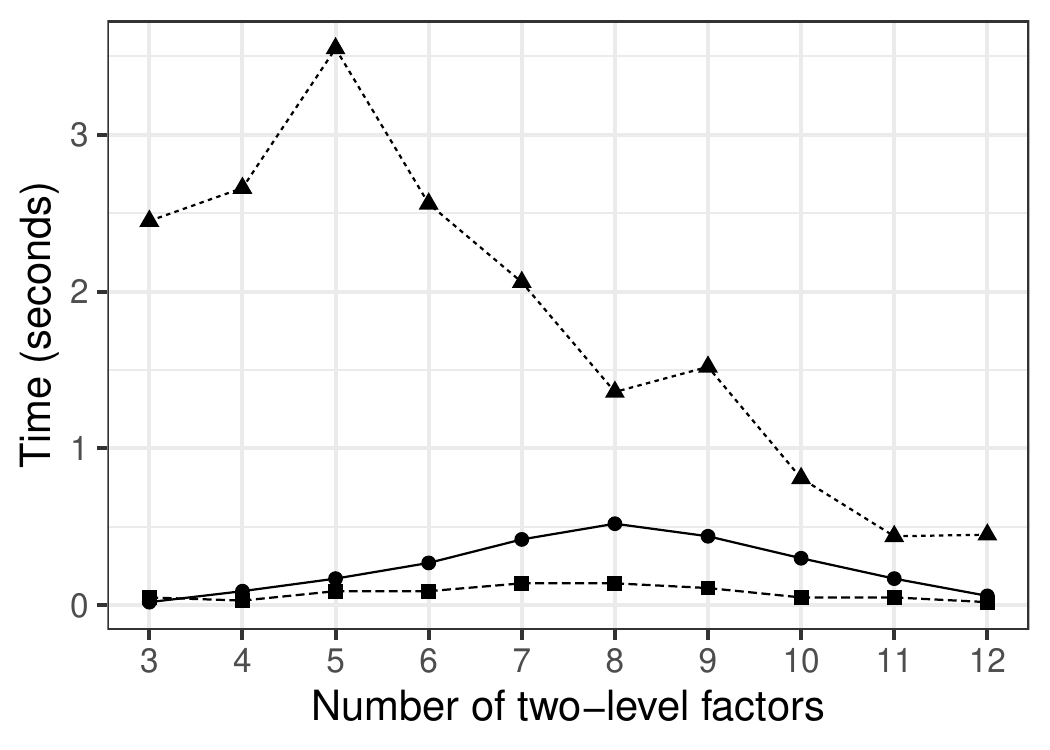}%
  \label{fig:test_3}%
}

\caption{Enumeration times for 32-run resolution \RNum{3} $4^12^{n-p}$ designs, 64-run resolution \RNum{4} $4^12^{n-p}$ designs and 64-run resolution \RNum{4} $4^22^{n-p}$ designs, where $n$ represents the number of two-level factors. Methods: ST-NAUTY $(\sqbullet)$, DOP-NAUTY $(\bullet)$, MCS-regular $(\blacktriangle)$}
\label{fig:test_cases}
\end{figure}

We apply the DOP-NAUTY, ST-NAUTY and MCS-regular methods to three test cases: (a) all 32-run designs of resolution 
\RNum{3} with one four-level factor, (b) all 64-run designs of resolution 
\RNum{4} with one four-level factor, and (c) all 64-run designs of resolution 
\RNum{4} with two four-level factors.
Figure \ref{fig:test_cases} shows the enumeration times for the three methods applied to the three test cases.
For all test cases, the MCS-regular method requires the longest computing time to enumerate all $4^m2^{n-p}$ designs.
The time difference between the DOP-NAUTY method (represented by black bullets) and the ST-NAUTY method (represented by black triangles) initially increases with the number of two-level factors.
From a certain number of two-level factors onwards, the computing time difference starts to drop again due to the scarcity of $4^m2^n$ designs with many two-level factors.
In all test cases considered and for each number of two-level factors, the ST-NAUTY algorithm is the fastest of the three for the enumeration of $4^m2^{n-p}$ designs so that we use this method to generate our catalog.

\subsection{Catalog}
\label{sec:catalog}
Since we only construct four-level factors from basic factors, and we need two basic factors to construct a single four-level factor, we consider a maximum of $\lfloor k / 2 \rfloor$ four-level factors for a design involving $2^{k}$ runs.
Using the ST-NAUTY algorithm, we completely enumerate all designs for the cases presented in Table \ref{tab:enumeration}.

\begin{table}[htbp]
  \centering
  \caption{All cases considered for the complete enumeration. For each case, we enumerate all $4^m2^{n-p}$ designs of a specific run size, with a resolution of at least $R$, and up to Max. $n$ two-level factors.}
    \begin{tabular}{lrrr}
    \toprule
    \multicolumn{1}{c}{Run size} & \multicolumn{1}{c}{$m$} & \multicolumn{1}{c}{$R$} & \multicolumn{1}{c}{Max. $n$} \\
    \midrule
    16 & 1 & \RNum{3} & 12 \\
    16 & 2 & \RNum{3} & 9 \\
    32 & 1,2 & \RNum{3} & 20 \\
    64 & 1 & \RNum{3} & 16 \\
    64 & 2 & \RNum{3} & 12 \\
    64 & 3 & \RNum{3} & 9 \\
    64 & 1,2,3 & \RNum{4} & 20 \\
    128 & 1,2,3 & \RNum{4} & 20 \\
    \bottomrule
    \end{tabular}%
  \label{tab:enumeration}%
\end{table}%

For 16-run designs, it is impossible to generate regular designs with one four-level factor and more than 12 two-level factors or designs with two four-level factors and more than nine two-level factors.
For the enumeration of 64-run designs with 1, 2, or 3 four-level factors, a resolution of at least \RNum{3}, and 17-20, 13-20, and 10-20 two-level factors, respectively, we use a bounded enumeration, as explained in Section 4.3, because the resolution-\RNum{3} designs for these numbers of two-level factors are too numerous.\\

Table \ref{tab:32-64-enumeration} presents the numbers of non-isomorphic designs with a resolution of at least \RNum{3} for run sizes of 16, 32, and 64.
Designs enumerated using the bounded enumeration are indicated with an asterisk.
While \citeauthor{ankenman_design_1999}'s \citeyear{ankenman_design_1999} catalog provided one design for each number of factors, we present all non-isomorphic 16-run and 32-run designs.
We also tackle resolution-\RNum{3} 64-run designs and found more than 33 million designs, all of which are new.
Table \ref{tab:128-256-enumeration} presents the numbers of non-isomorphic designs with a resolution of at least \RNum{4} for run sizes of 64 and 128.
With this table, we present more than 6 million new designs.
This offers the prospect of being able to tackle many future design-of-experiments problems in a systematic rather than an ad hoc fashion.\\

The catalog can be explored online through an interactive web app at \url{https://abohyndoe.shinyapps.io/fatldesign-selection-tool/}.
In the app, filters are provided to search the catalog based on run size, resolution, number of four-level factors, and number of two-level factors.
For the selected designs, the columns needed to generate the design matrices, the complete word length patterns, the $\mathbf{A}_{3}^{m}$ vectors, the $\mathbf{A}_{4}^{m}$ vectors, and the $\mathbf{A}_{5}^{m}$ vectors are shown.
All designs are also available from the authors.

\begin{table}[hbtp]
	\centering
	\caption{Numbers of non-isomorphic $4^m2^{n-p}$ designs  of resolution \RNum{3} with $N=16$, $N=32$ and $N=64$ runs.} 
	\label{tab:32-64-enumeration}
	\begin{tabular}{rrrrrrrr}
		\toprule
		\multirow{2}{*}{$n$} & \multicolumn{2}{c}{$N=16$} & \multicolumn{2}{c}{$N=32$} & \multicolumn{3}{c}{$N=64$}\\
		\cmidrule{2-8}
			& $m=1$ & $m=2$ & $m=1$ & $m=2$ & $m=1$ & $m=2$ & $m=3$\\
		\midrule
		1 	& - & 1 &  	-	&   1 	&  	-		&  	-					&   2\phantom{*} 		\\ 
		2 	& 1 & 2 &  	-	&   3 	&  	-		&   1\phantom{*} 		&  10\phantom{*} 		\\ 
		3 	& 3 & 4 &   1 	&  11 	&  	-		&   6\phantom{*}		&  64\phantom{*} 		\\ 
		4 	& 5 & 5 &   5 	&  38 	&   1 		&  33\phantom{*} 		& 453\phantom{*} 		\\ 
		5 	& 7 & 5 &  14 	& 109 	&   7 		& 178\phantom{*} 		& 3315\phantom{*} 		\\ 
		6 	& 9 & 4 &  37 	& 285 	&  31 		& 944\phantom{*} 		& 23290\phantom{*} 		\\ 
		7 	& 7 & 2 &  82 	& 650 	& 120 		& 4755\phantom{*} 		& 148180\phantom{*} 	\\ 
		8 	& 6 & 1 & 159 	& 1307 	& 449 		& 23133\phantom{*} 		& 838847\phantom{*} 	\\ 
		9 	& 4 & 1 & 285 	& 2307 	& 1619 		& 106841\phantom{*} 	& 4206184\phantom{*} 	\\ 
		10 	& 2 & 0 & 462 	& 3535 	& 5717 		& 462544\phantom{*} 	& 435436*				\\ 
		11 	& 1 & 0 & 669 	& 4697 	& 19776 	& 1854971\phantom{*}	& 362518*				\\ 
		12 	& 1 & 0 & 888 	& 5423 	& 66755 	& 6842015\phantom{*}	& 274634*				\\ 
		13 	& 0 & 0 & 1047 	& 5423 	& 216678 	& 7415*					& 173318*				\\ 
		14 	& 0 & 0 & 1106 	& 4697 	& 668229 	& 3098*					& 100340*				\\ 
		15 	& 0 & 0 & 1047 	& 3535 	& 1938759 	& 1193*					& 50380*				\\ 
		16 	& 0 & 0 & 889 	& 2308 	& 5258438	& 560*					& 21093*				\\ 
		17 	& 0 & 0 & 670 	& 1308 	& 3105515*	& 238*					& 6844*					\\ 
		18 	& 0 & 0 & 464 	& 652 	& 3398532*	& 124*					& 1634*					\\ 
		19 	& 0 & 0 & 289 	& 289 	& 2401793*	& 44*					& 247*					\\ 
		20 	& 0 & 0 & 165 	& 114 	& 837086*	& 11*					& 16*					\\ 
		\bottomrule
		\multicolumn{6}{l}{\footnotesize *Number of designs obtained using bounded enumeration}\\
	\end{tabular}
\end{table}

\begin{table}[hbtp]
	\centering
	\caption{Numbers of non-isomorphic $4^m2^{n-p}$ designs  of resolution $IV$ with $N=64$ and $N=128$ runs.} 
	\label{tab:128-256-enumeration}
	\begin{tabular}{rrrrrrr}
		\toprule
		\multirow{2}{*}{$n$} & \multicolumn{3}{c}{$N=64$} & \multicolumn{3}{c}{$N=128$} \\
		\cmidrule{2-7}
		& $m=1$ & $m=2$ & $m=3$ & $m=1$ & $m=2$ & $m=3$ \\
		\midrule
		1 	&  	-	&  	-	&   1 	&  	-		&  	-		&   1 		\\ 
		2 	&  	-	&  	-	&   2 	&  	-		&  	-		&   3 		\\ 
		3 	&  	-	&   3 	&   4 	&  	-		&   1 		&  13 		\\ 
		4 	&   1 	&   7 	&   7 	&  	-		&   6 		&  67 		\\ 
		5 	&  	5 	&  13 	&   7 	&   1 		&  24 		& 360 		\\ 
		6 	&  10 	&  25 	&   5	&   7 		& 102 		& 1967 		\\ 
		7 	&  17 	&  29 	&  	2	&  24 		& 438 		& 9451 		\\ 
		8 	&  32 	&  28 	&  	0	&  76 		& 1880 		& 36124 	\\ 
		9 	&  41 	&  17 	&  	0	& 263 		& 7501 		& 104631 	\\ 
		10 	&  43 	&   9 	&  	0	& 895 		& 26212 	& 226693 	\\ 
		11 	&  40 	&   4 	&  	0	& 2851 		& 74969 	& 366113 	\\ 
		12 	&  29 	&  	2	&  	0	& 8640 		& 171266 	& 442617 	\\ 
		13 	&  17 	&  	0	&  	0	& 23210 	& 308495 	& 404157 	\\ 
		14 	&  11 	&  	0	&  	0	& 53458 	& 440218 	& 283471 	\\ 
		15 	&  	8	&  	0	&	0	& 103531 	& 503159 	& 156987 	\\ 
		16 	&  	0	&  	0	&  	0	& 168157 	& 470362 	& 70809 	\\ 
		17 	&  	0	&  	0	&  	0	& 231345 	& 368468 	& 26810 	\\ 
		18 	&  	0	&  	0	&  	0	& 275786 	& 247266 	& 8625		\\ 
		19 	&  	0	&  	0	&  	0	& 292509 	& 143885 	& 2408		\\ 
		20 	&  	0	&  	0	&  	0	& 281548 	& 73023 	& 604 		\\ 
		\bottomrule
	\end{tabular}
\end{table}

\subsection{Bounded enumeration}
\label{sec:bound}
Even with our efficient ST-NAUTY algorithm, enumerating all non-isomorphic $4^m2^{n-p}$ designs becomes computationally prohibitive for large values of $n$.
This proves to be the case with our enumeration of 64-run designs of resolution \RNum{3}, where, for certain numbers of two-level factors, more than four million non-isomorphic designs are enumerated.
From that point onwards, we extend designs using an upper bound  for values contained within the $\mathbf{A}_{3}$ vector in order to reduce the number of candidate designs generated.
We define the bound as $\boldsymbol{\delta}^{3}_{m,n} = \left(\delta^{3}_{m,n,0},\ldots,\delta^{3}_{m,n,m}\right)$, where $\delta^{3}_{m,n,t}$ is the upper bound on the number of words of length 3 and type $t$ for $4^{m}2^{n-p}$ designs.
More specifically, a candidate $4^{m}2^{(n+1)-(p+1)}$ design of resolution \RNum{3} with $\mathbf{A}_3 = \left(A_{30},\ldots,A_{3m}\right)$ is only added to the complete set $\Tilde{C}_{m,n+1,p+1}$ and submitted to the reduction step with NAUTY if $A_{3t} \leq \delta^3_{m,n+1,t}$ for all $t$ from $0$ to $m$.\\

We enumerate designs with up to 20 two-level factors, so we define the $\boldsymbol{\delta}^{3}_{m,n}$ values for $n \leq 20$, as follows.
We construct $4^{m}2^{20-p}$ designs from all $2^{20 + 2m}$ designs from \citet{chen_catalogue_1993}.
Let $D_{m,20}$ be the minimum aberration design of type $m$ among all the $4^m2^{20-p}$ designs thus created.
We use the $\mathbf{A}_{3}$ vector of $D_{m,20}$ to define $\boldsymbol{\delta}^{3}_{m,20}$.
Next, we consider all 20 delete-one-factor projections (DOPs) of $D_{m,20}$, and take the one with the worst $\mathbf{A}_3$ vector.
The worst $\mathbf{A}_3$ vector is defined as the one that sequentially maximizes $A_{3t}$ for $t$ going from $m$ to zero.
This worst $\mathbf{A}_3$ defines $\boldsymbol{\delta}^{3}_{m,19}$.
So, we do not extend any $4^m 2^{19-(p-1)}$ design with $A_{3t}>\delta^{3}_{m,19,t}$ for $t=0,\ldots,m$. 
If there is more than one design with the worst $\mathbf{A}_3$ vector, we perform an isomorphism check and we keep the non-isomorphic designs with the worst $\mathbf{A}_3$ vector in a set denoted by $S_{m,19}$.
For the bound on $4^m2^{18-(p-2)}$ designs, we then consider all DOPs from all $4^m 2^{19-(p-1)}$ designs in $S_{m,19}$.
The procedure continues until we reach a value of $n$ for which the enumeration was complete.\\

There are two advantages to this bound definition.
First, we ensure that the bound will never be so restrictive that no designs are enumerated.
Second, we ensure that the enumerated designs are at least as good in terms of type-$m$ aberration as the ones that we constructed as a reference to create the bound.
By focusing on aberration of type $m$, we favor designs that minimize the aliasing involving the four-level factors, which cannot be easily constructed from existing two-level designs.\\

Table \ref{tab:bound_values} provides the bound values for $4^m2^{n-p}$ designs of resolution \RNum{3} with 64 runs and $m=1,2$ and $3$, whenever the bounded enumeration was used.
In a supplementary table, we present details on the number of candidate designs that are rejected prior to the isomorphism reduction step because at least one entry of their $\mathbf{A}_3$ vector exceeds the bound.

\begin{table}[htbp]
  \centering
  \caption{$\boldsymbol{\delta}^{3}_{m,n}$ bounds on the $\mathbf{A}_3$ vectors for $4^m2^{n-p}$ designs of resolution \RNum{3} and run size 64 given as $(\delta_{m,0},\ldots,\delta_{m,m})$. Extension steps for which the bounded enumeration was not used are indicated with a dash (-).}
    \begin{tabular}{rrrr}
    \toprule
    \multicolumn{1}{c}{$n$} & \multicolumn{1}{c}{$m=1$} & \multicolumn{1}{c}{$m=2$} & \multicolumn{1}{c}{$m=3$} \\
    \midrule
    20 &  (0, 5) &  (0, 16, 0) &  (0, 22, 6, 0) \\
    19 &  (0, 5) &  (0, 16, 0) &  (0, 21, 6, 0) \\
    18 &  (0, 5) &  (0, 16, 0) &  (0, 20, 6, 0) \\
    17 &  (0, 5) &  (0, 16, 0) &  (0, 19, 6, 0) \\
    16 & - &  (0, 16, 0) &  (0, 18, 6, 0) \\
    15 & - &  (0, 14, 0) &  (0, 16, 6, 0) \\
    14 & - &  (0, 13, 0) &  (0, 14, 6, 0) \\
    13 & - &  (0, 12, 0) &  (0, 11, 6, 0) \\
    12 & - & - &  (0, 9, 6, 0) \\
    11 & - & - &  (0, 7, 6, 0) \\
    10 & - & - &  (0, 6, 6, 0) \\
    9 & - & - & - \\
    \bottomrule
    \end{tabular}%
  \label{tab:bound_values}%
\end{table}%

\section{Applications}
\label{sec:application}
In this section, we revisit the two motivating examples for this paper.
More specifically, we investigate whether the ad hoc designs used  can be improved upon by one or more designs in our catalog.

\subsection{Chemical etching experiment}
The first practical case presented by \cite{schoen_designing_1999} is a chemical etching experiment involving 32 runs with two four-level factors and 12 two-level factors.
In this experiment, a run involves chemical synthesis of a catalyst performed on a gauze in an autoclave.
Since the synthesis takes time to prepare and to run, the 32 runs were divided over eight days. Four of the two-level factors were varied between the days, while the remaining eight two-level factors as well as the two four-level factors were varied within the days. Therefore, the experiment had a split-plot structure with days as whole plots and syntheses as sub-plots.\\

Table \ref{tab:32-64-enumeration} shows that there are 5423 non-isomorphic $4^{2}2^{12-11}$ designs.
We made a preliminary inventory of these designs in terms of the word length patterns of type 0 and type 2.
Tables \ref{tab:N32_aberration_type0} and \ref{tab:N32_aberration_type2} show the number of words of length 3 and length 4 of the best $4^22^{12-11}$ designs in terms of aberration of type 0 and aberration of type 2, respectively.
The split-plot structure of the experiment requires four two-level factors to be varied between eight whole plots, defined by the eight days.
This means that, for the whole plots, a $2^{4-1}$ design is needed.
Therefore, any $4^{2}2^{12-11}$ design with at least one word of length 4 and type 0 can be used for the experiment.
All designs presented in Tables \ref{tab:N32_aberration_type0} and \ref{tab:N32_aberration_type2} have a least 38 of these words so that each of them can accommodate the split-plot structure of the experiment.
It turns out that the design used by \citet{schoen_designing_1999} for the experiment has minimum aberration of type 0; it is design 5381, the first one in Table \ref{tab:N32_aberration_type0}.
The design has no words of length 3 and type 0, meaning that no main effects of a two-level factor are aliased with a two-factor interaction involving other two-level factors, and 10 words of length 3 and type 1, meaning that 10 two-factor interactions involving two-level factors are aliased with a main effect of a four-level factor.
The design also has four words of length 3 and type 2, so that the main effects of four two-level factors are aliased with an interaction between the four-level factors.\\

\begin{table}[htbp]
  \centering
  \caption{Number of words of length 3 and length 4 of the best 32-run $4^{2}2^{12-11}$ designs according to the aberration of type 0.}
    \begin{tabular}{cccccccc}
    \toprule
    Rank & ID & $A_{30}$ & $A_{31}$ & $A_{32}$ & $A_{40}$ & $A_{41}$ & $A_{42}$\\
    \midrule
    1 & 5381 & 0 & 10 & 4 & 38 & 68 & 24\\
    2 & 4902 & 0 & 17 & 6 & 38 & 34 & 13\\
    3 & 3633 & 0 & 18 & 5 & 38 & 34 & 13\\
    4 & 3632 & 0 & 18 & 6 & 38 & 34 & 12\\
    5 & 4901 & 0 & 18 & 6 & 39 & 32 & 12\\
    \bottomrule
    \end{tabular}
  \label{tab:N32_aberration_type0}
\end{table}

The design used by \citet{schoen_designing_1999} does not have minimum aberration of type 2.
More specifically, there are 1356 designs that are better in terms of aberration of type 2.
For this reason, it is not shown in Table \ref{tab:N32_aberration_type2}.
It has four words of length 3 and type 2, while the design with minimum aberration of type 2 has no such words.
The absence of words of length 3 and type 2 is a useful feature if the four-level factors are likely to interact with each other because the main effects of the two-level factors are then not aliased with the interaction among the four-level factors.
This demonstrates the utility of having a catalog of $4^22^{12-11}$ designs at hand.

\begin{table}[htbp]
  \centering
  \caption{Number of words of length 3 and length 4 of the best 32-run $4^{2}2^{12-11}$ designs according to the aberration of type 2.}
    \begin{tabular}{cccccccc}
    \toprule
    Rank & ID & $A_{32}$ & $A_{31}$ & $A_{30}$ & $A_{42}$ & $A_{41}$ & $A_{40}$\\
    \midrule
    1 & 3346 & 0 & 24 & 0 & 42 & 0 & 39 \\
    2 & 1286 & 0 & 25 & 0 & 41 & 0 & 38 \\
    3 & 159 & 0 & 26 & 0 & 40 & 0 & 38 \\
    4 & 162 & 0 & 26 & 0 & 40 & 0 & 39 \\
    5 & 161 & 0 & 27 & 0 & 39 & 0 & 38 \\
    \bottomrule
    \end{tabular}
  \label{tab:N32_aberration_type2}
\end{table}

\subsection{Cheese-making experiment}
The second practical case presented by \cite{schoen_designing_1999} is a 10-factor cheese-making experiment involving 128 runs with one four-level factor and nine two-level factors.
In this experiment, milk was distributed over several curds production tanks.
In each of these tanks, the curds were formed and transported to presses forming the individual cheeses.
Seven factors were varied over the curds production tanks, while three factors were varied at the level of the individual cheeses.
Due to various practical limitations, the experiment had to be conducted using a split-plot experimental design involving 32 whole plots, corresponding to the curds productions, each containing four sub-plots corresponding to the individual cheeses.\\

Due to our complete enumeration, we have a catalog of all non-isomorphic $4^12^{9-4}$ designs involving 128 runs.
Table \ref{tab:128-256-enumeration} shows that there are 263 such designs.
We evaluated each of these designs in terms of the word length pattern of type 0, and the word length pattern of type 1.
Table \ref{tab:ma_128_designs_type1} shows the number of words of length 4 and length 5, of the five best 128-run $4^{1}2^{9-4}$ designs, according to the aberration of type 1.
\cite{schoen_designing_1999} actually used design 222, the third best, with $A_{40}=1, A_{41}=0, A_{50}=2$ and $A_{51}=5$.
The split-plot structure of the experiment required seven whole-plot factors to be studied over 32 curds productions.
This means that for the whole plots, a $2^{7-2}$ design was needed, which can at best have a resolution of \RNum{4} \citep{chen_catalogue_1993}.
The minimum aberration 32-run design with seven two-level factors has $A_4=1$ and $A_5=2$.
So, the cheese-making design must have at least one word of length 4 and type 0, and at least two words of length 5 and type 0.
This is incompatible with designs 262, 263, and 230 in Table \ref{tab:ma_128_designs_type1}.
The design that was used by \citet{schoen_designing_1999} thus has the best aberration of type 1, given the split-plot structure.\\

\begin{table}[htbp]
	\centering
	\caption{Number of words of length 4 and length 5 of the best 128-run $4^12^{9-4}$ designs
	according to the aberration of type 1.} 
		\begin{tabular}{cccccc}
			\toprule
			Rank & ID &$A_{41}$ & $A_{40}$ & $A_{51}$ & $A_{50}$ \\
			\midrule
			1 & 262 & 0 & 0 & 6 & 2 \\
			2 & 263 & 0 & 0 & 9 & 0 \\
			3 & 222 & 0 & 1 & 5 & 2 \\
			4 & 230 & 0 & 1 & 6 & 1 \\
			5 & 223 & 0 & 1 & 6 & 2 \\
			\bottomrule
		\end{tabular}
	\label{tab:ma_128_designs_type1}
\end{table}

\citeauthor{schoen_designing_1999}'s design is the 38th best in terms of aberration of type 0, among the 263 $4^12^{9-4}$ designs in the catalog. 
However, none of the 37 better designs can accommodate the split-plot structure because none of them have words of length 4 and type 0 and words of length 5 and type 0.
We conclude that the design used in the cheese-making experiment has the smallest aberration both of type 0 and type 1, given the split-plot structure.

\subsection{Sensor fields simulation experiment}
\citet{katic2011optimization} discussed a sensor fields simulation experiment involving 32 runs, two four-level factors, and five two-level factors.
The experiment was intended to determine the ideal position for one or more undersea sensors given the environmental conditions of the sea, the features of the sensor, and the characteristics of the object entering the sensor field.
In the experiment, \citet{katic2011optimization} want to quantify how the different inputs in the program affect the placement of the sensors.
The initial position and direction of an object entering the sensor field are both modeled using a probability density function (PDF) that can take four different forms.
For this reason, two four-level factors were used to represent the four possible PDFs for the initial position and direction.
The five two-level factors represent other features of the sensors, and of the object entering the sensor field.\\

To generate the design, the author started with the minimum aberration $2^{9-4}$ design and created two four-level factors, $A=(b,c)$ and $B=(d,e)$, using the grouping scheme of \cite{wu_construction_1989}.
The final design obtained is a 32-run $4^{2}2^{5-4}$ design with $\mathbf{A}^{0}_3=(0,1,1)$, $\mathbf{A}^{0}_4=(0,4,5)$, and $\mathbf{A}^{0}_5=(0,1,2)$.
Table \ref{tab:32-64-enumeration} shows that there are 109 non-isomorphic $4^{2}2^{5-4}$ designs with 32 runs.
Table \ref{tab:ex3_N32_aberration_type0} shows the number of words of lengths 3 and 4 for the five best $4^{2}2^{5-4}$ designs in terms of aberration of type 0.
We see that the design used in this experiment is design 62 and that it is only the fourth best in terms of aberration of type 0.
\citet{wu_minimum_1993} and \citet{ankenman_design_1999} both provide a design with minimum aberration of type 0 so that \citet{katic2011optimization} could have used this design for their research.
It is isomorphic to design 109 in our catalog. 

\begin{table}[htbp]
  \centering
  \caption{Number of words of length 3 and length 4 of the best 32-run $4^22^{5-4}$ designs
  according to the aberration of type 0.}
    \begin{tabular}{cccccccc}
    \toprule
    Rank  & ID  & $A_{30}$  & $A_{31}$  & $A_{32}$  & $A_{40}$  & $A_{41}$  & $A_{42}$ \\
    \midrule
     1  & 109 & 0         & 0         & 1         & 1         & 4         & 6 \\
     2  & 106 & 0         & 0         & 2         & 1         & 4         & 4 \\
     3  & 107 & 0         & 0         & 2         & 0         & 4         & 4 \\
     4  & 62  & 0         & 1         & 1         & 0         & 4         & 5 \\
     5  & 79  & 0         & 1         & 1         & 0         & 3         & 5 \\
    \bottomrule
    \end{tabular}%
  \label{tab:ex3_N32_aberration_type0}%
\end{table}%

Minimizing aberration of type 0 minimizes the confounding between main effects and two-factor interactions involving two-level factors. 
However, if it is likely that the four possible PDFs for the initial position and direction interact, it could be useful to minimize the confounding of the main effects of the two-level factors with the interaction among the four-level factors.
This is achieved by minimizing the aberration of type 2. 
To look for designs that minimize the aberration of type 2, we evaluated the 109 non-isomorphic $4^{2}2^{5-4}$ designs with 32 runs in terms of the word length pattern of type 2.
They are presented in Table \ref{tab:ex3_N32_aberration_type2}.
Design 80 has minimum aberration of type 2, with $\mathbf{A}^{2}_3=(0,2,0)$ and $\mathbf{A}^{2}_4=(0,0,8)$, while design 62, i.e., the one used by \citet{katic2011optimization}, ranks 16th in terms of aberration of type 2.
As design 62, design 80 has no words of length 3 and type 0, but it has 8 words of length 4 instead of 9.
In conclusion, designs 109 or design 80, are better than design 62 in terms of aberration.
In any case, both options can be found in our catalog.

\begin{table}[htbp]
  \centering
  \caption{Number of words of length 3 and length 4 of the best 32-run $4^22^{5-4}$ designs
  according to the aberration of type 2.}
    \begin{tabular}{cccccccc}
    \toprule
    Rank  & ID  & $A_{32}$  & $A_{31}$  & $A_{30}$  & $A_{42}$  & $A_{41}$  & $A_{40}$ \\
    \midrule
    1 & 80 & 0 & 2 & 0 & 8 & 0 & 0 \\
    2 & 43 & 0 & 2 & 0 & 8 & 0 & 1 \\
    3 & 41 & 0 & 3 & 0 & 7 & 0 & 0 \\
    4 & 42 & 0 & 3 & 0 & 7 & 0 & 1 \\
    5 & 10 & 0 & 4 & 0 & 6 & 0 & 0 \\
    \bottomrule
    \end{tabular}%
  \label{tab:ex3_N32_aberration_type2}
\end{table}%

\section{Discussion}
\label{sec:discussion}

In this paper, we developed an efficient enumeration procedure for regular $4^m2^{n-p}$ designs.
Using this procedure, we created an extensive catalog of such designs for run sizes of 16, 32, 64, and 128, with one, two, or three four-level factors and up to 20 two-level factors.
For a few cases, we could not enumerate all non-isomorphic designs.
However, for all of these cases, we were able to provide a set of good designs rather than just the minimum aberration designs.
To do so, we implemented a bounded enumeration that is restrictive enough for the enumeration to be computationally feasible while still providing a good subset of all possible designs.
The entire catalog we obtained is a major addition to the currently available catalogs for $4^m2^{n-p}$ designs \citep{wu_minimum_1993,ankenman_design_1999}.
Furthermore, we built an interactive web application that allows anyone to browse through the catalog and find a design based on several criteria.\\

The catalog opens up possibilities for future work.
As several millions of designs are now available, a precise characterization of these designs would be helpful.
The possibility of optimizing the designs according to multiple criteria could help researchers better choose the design that fits their experimental needs.
Looking back at the cheese-making example, the main argument for the choice of the design was the split-plot layout imposed by the experimental conditions.
So, interesting follow-up work would be to characterize the designs' potential to be run as a split-plot design.\\

Another restriction in the randomization, often seen in screening experiments, is when designs have to be blocked.
In some cases, there is one blocking factor  \citep{sun_optimal_1997,sartonoBlockingOrthogonalDesigns2015,schoenOrthogonalBlockingArrangements2019}, while, in other cases, there are two blocking factors \citep{godolphin_construction_2019,vo-thanhRowcolumnArrangementsRegular2020}.
Thus, it would also be interesting to study the different blocking schemes that could be applied to our newly enumerated designs.

\bigskip
\if0\blind
{
\begin{center}
{\large\bf ACKNOWLEDGMENTS}
\end{center}
We would like to thank Hongquan Xu as well as Kenneth Ryan for providing us with the code from their respective papers and taking the
time to answer our questions. The authors would also like to thank the anonymous referees for improving this paper with their constructive
comments.
} \fi

\begin{center}
{\large\bf CONFLICT OF INTEREST}
\end{center}
None declared

\begin{center}
{\large\bf FUNDING}
\end{center}
This research was funded by the FWO.

\begin{center}
    {\large\bf DATA AVAILABILITY}
\end{center}
The designs are accessible through a web app at \url{https://abohyndoe.shinyapps.io/fatldesign-selection-tool/}.
The code to reproduce the methodology of the paper is available online at the GitHub repository \url{https://github.com/ABohynDOE/enumeration_fatld}.

\begin{center}
{\large\bf SUPPLEMENTARY MATERIAL}
\end{center}

\begin{description}
\item[pseudocode.pdf:] Pseudo-code schemes for the ST-NAUTY and DOP-NAUTY methods.
\end{description}

\begin{appendices}
\section{Proofs}
\label{appendix:proofs}

\begin{lemma}
\label{lem:first_lemma}
Consider a $4^{m}2^{(n+1)-(p+1)}$ design $D$ involving $N$ runs.
Let $D_{(i)}$ be the ${4^{m}2}^{n-p}$ subdesign obtained by deleting the $i$th two-level column of $D$.
If $D_{(i)}$ has minimum aberration among all the subdesigns of $D$ then the $i$th column is a product of some two-level columns and thus $D_{(i)}$ has $N$ distinct runs.
\end{lemma}

\begin{proof}
Suppose that the result is not true. 
Then, the $i$th column is independent of the other columns and thus does not appear in the defining relation of $D$.
In that case, we can choose another column that does appear in the defining relation; deleting that other column would yield a design having less aberration than $D_{(i)}$, which is a contradiction.
\end{proof}

\begin{lemma}
\label{lem:second_lemma}
For any $4^{m}2^{n-p}$ design $D$ in the minimum complete set $C^{R}_{m,n,p}$, adding a two-level column yields a $4^{m}2^{(n+1)-(p+1)}$ candidate design $D_c$. 
Such a candidate can be discarded if its resolution is lower than $R$ or if $D$ does not have MA among all DOPs of $D_c$.
The resulting set, $\widetilde{C}^{R}_{m,n+1,p+1}$, is a complete set.
\end{lemma}

\begin{proof}
For a $4^{m}2^{n-p}$ design $D$ involving $2^{k}$ runs, the $m$ four-level factors can be decomposed into $m$ mutually exclusive triplets of two-level pseudo-factors of the form $(a_1,a_2,a_3)$, where $a_3=a_1 a_2$.
Therefore, $D$ involves $3m + n \leq 2^{k} - 1$ two-level factors.
We restrict our attention to the case where the four-level subdesign is a full factorial design that is possibly replicated, that is when $2m \leq k$.
In that case, each triplet of pseudo-factors consists of a pair of basic factors and their Hadamard product. 
As a result, there are $k^{\prime} = k - 2m$ further basic factors in the design.\\

If $n \leq k^{\prime}$, then the basic factors not involved in the definition of the four-level factors are used to define additional two-level factors. 
If $n > k^{\prime}$, we show by induction that every possible $4^{m}2^{n-p}$ design in $2^{k}$ runs is isomorphic to a design in $C_{m,n,p}$ obtained with the DOP procedure.

That this is true for $n = k^{\prime} + 1$ is trivial, since the parent design is a full factorial design. 
Any additional column is then a product of basic factors, and the parent design must have MA among all DOPs.
Now, suppose that Lemma \ref{lem:second_lemma} is true for $n = k^{\prime}  + l$.
Consider $n + 1 = k^{\prime} + l + 1$.
Let $D_{c} = ({C_{1},\ldots,C_{m},c}_{1},\ldots,\ c_{n + 1})$ be a $4^{m}2^{(n+1)-(p+1)}\ $ candidate design involving $N$ runs where $C_i$ is the $i$th four-level factor of the design and $c_i$ is the $i$th two-level factor of the design.
Suppose that $D_{c(n+1)}$ has MA among all DOPs of $D_{c}$.
Lemma \ref{lem:first_lemma} implies that $D_{c(n+1)}$ has $N$ distinct runs.
By the assumption for $n = k^{\prime}  + l$, there exists a $4^m2^{n-p}$ design $D_n$ in $C_{m,n,p}$ that is isomorphic to $D_{c(n+1)}$.
Let $\kappa,\rho$ and $\sigma$ be the column permutations, row permutations and sign switches, respectively, that form the isomorphic map $\pi$ from $D_{c(n+1)}$ to $D_n$, that is, $D_n = \pi(D_{c(n+1)})$.
Out of the three operations in $\pi$, only $\rho$ can be applied to the last column, so let $\rho(c_{n+1})$ be the result of the row permutations $\rho$ applied to $c_{n+1}$.
Now, by applying $\pi$ to $D_c$ we obtain the following result: $\pi(D_c) = \pi\left(D_{c(n+1)},c_{n+1}\right) = \left(\pi\left(D_{c(n+1)}\right),\pi\left(c_{n+1}\right)\right) = \left( D_n, \rho(c_{n+1})\right)$.
By definition, $D_n$ is in $C_{m,n,k}$.
By Lemma A.1, column $c_{n+1}$ of $D_c$ is a product of some other columns of $D_c$ and since all columns are considered in the extension procedure, $\rho(c_{n+1})$ is entertained in the extension procedure.
Therefore, $\pi(D_c)$ is entertained in this modified construction procedure.
This means that there is an isomorphic map from $D_c$ to a design in $C_{m,n+1,p+1}$ and that $D_c$ is isomorphic to a design in $C_{m,n + 1,p+1}$.
Observing that this reasoning can be applied to any value of $n$ and
$l$ and that it is true for $n = k^{\prime} + 1$ completes the proof.
\end{proof} 

\end{appendices}

\bibliographystyle{apalike} 
\bibliography{references} 


\end{document}